\documentstyle[12pt]{article}
\begin{document}
\newcommand{\eqn}[1]{eq.(\ref{#1})}
\newcommand{\Eqn}[1]{Eq.(\ref{#1})}

\renewcommand{\section}[1]{\addtocounter{section}{1}
\vspace{5mm} \par \noindent
  {\bf \thesection . #1}\setcounter{subsection}{0}
  \par
   \vspace{2mm} } 
\newcommand{\sectionsub}[1]{\addtocounter{section}{1}
\vspace{5mm} \par \noindent
  {\bf \thesection . #1}\setcounter{subsection}{0}\par}
\renewcommand{\subsection}[1]{\addtocounter{subsection}{1}
\vspace{2.5mm}\par\noindent {\em \thesubsection . #1}\par
 \vspace{0.5mm} }
\renewcommand{\thebibliography}[1]{ {\vspace{5mm}\par \noindent{\bf
References}\par \vspace{2mm}}
\list
 {\arabic{enumi}.}{\settowidth\labelwidth{[#1]}\leftmargin\labelwidth
 \advance\leftmargin\labelsep\addtolength{\topsep}{-4em}
 \usecounter{enumi}}
 \def\newblock{\hskip .11em plus .33em minus .07em}
 \sloppy\clubpenalty4000\widowpenalty4000
 \sfcode`\.=1000\relax \setlength{\itemsep}{-0.4em} }

\def\Buildrel#1\under#2{\mathrel{\mathop{#2}\limits_{#1}}}
\def\pro#1{\!\Buildrel
	\raise 4pt\hbox{\the\scriptscriptfont0 #1}\under\circ\!}
\def\Pro#1{\!\Buildrel#1\under\circ\!}
\def\half{{1\over 2}}

\begin{flushright}
HUB-IEP 95/31 \\ hep-th/9512134 \\ December 1995
\end{flushright}

\vspace{4mm}
\begin{center}
{\large \bf Octonions and Supersymmetry\footnote{
Talk presented at the Workshop on {\it
Gauge theories, applied supersymmetry and quantum
gravity}, Leuven, Belgium, July 10-14,
1995. } }
\vspace{1cm}

C.R.~PREITSCHOPF \\
\vspace{4mm}
{\em Institut f\"ur Physik}\\
{\em Humboldt-Universit\"at zu Berlin} \\
{\em Invalidenstr. 110, 10115 Berlin, Germany} \\
\end{center}
\centerline{ABSTRACT}
\vspace{0 mm}  
\begin{quote}\small
We apply the techniques of $S^7$-algebras to the
construction of N=5-8 superconformal algebras and
of S{\bf O}(1,9), a modification of SO(1,9) which
commutes with $S^7$-transformations. We discuss the
relevance of S{\bf O}(1,9) for off-shell super-Maxwell
theory in D=(1,9).
\end{quote}
\addtocounter{section}{1}
\par \noindent
  {\bf \thesection . N=5-8 Superconformal Algebras}
  \par
   \vspace{2mm} 

\noindent
We seek to generalize the $N=4$ superconformal algebra
\begin{eqnarray}
&J^I(z)J^J(w)\ =\ &{-c/3\over(z-w)^2}\  \delta^{IJ}\ +\ {1\over
		z-w}\ \epsilon^{IJK}\  \Big(J^K(z)+J^K(w)\Big) \ \ ,
							\nonumber\\
&J^I(z)G_a(w)\ =\ &{1\over z-w}\ \sigma_{ab}^I \ G_b(w)]\ \ ,
							\nonumber\\
&G_a(z)G_b(w)\ =\ &{  2c/3\over(z-w)^3}\  \delta_{ab}\
+\ {1\over(z-w)^2}\ \sigma_{ab}^I\ \Big( J^I(z)\ + \ J^I(w) \Big)  \
							\nonumber\\
&\phantom{ G_\alpha(z)G_\beta(w)\ =\    }
		&+\ {1\over z-w}\ \delta_{ab}\ \Big(L(z) + L(w)\Big)\ \
\label{N=4}
\end{eqnarray}
with
\begin{displaymath}
I,J,K \in \{1,2,3\} \qquad, a,b \in \{0,1,2,3\}
\end{displaymath}
and
\begin{displaymath}
\sigma_{ab}^I \ = \ [ e_a^* \ e_b \ e^{I*} ] .
\end{displaymath}
The symbols $e_a$ and $e^{I*}$ denote unit quaternions and their conjugates,
and by the bracket $[...]$ we mean the real part of a hypercomplex number.
The step from $N=4$ to $N>4$ proceeds by first generalizing SU(2) $\simeq$
$S^3$ to $S^7$. The latter is not a group manifold anymore, and therefore
we generalize the structure constants $\epsilon^{IJK}$ to structure
functions
\begin{equation}
T^{IJK}_X \ = \  [ (e^{I*} X^*) \ ( X e^J)  \ e^{K*} ]
\ =: \  [ (e^{I*} \Pro X  e^J)  \ e^{K*} ] \quad,
\label{torsion}
\end{equation}
defined
using imaginary unit octonions $e^I$ and the octonionic coordinate $X, |X|=1$
on the seven-sphere \cite{englert, berkon8, n8alg, cedpreit}:
\begin{eqnarray}
&J^I(z)J^J(w)\ =\ &{k\over(z-w)^2}\  \delta^{IJ}\ +\ {2\over
		z-w}\ :T^{IJK}_X(w)\  J^K(w): \ \ ,
							\nonumber\\
&J^I(z)X(w)\ =\ &{1\over z-w}\ X(w) \ e^{I*}\ \ ,
\label{S7}
\end{eqnarray}
where now of course $I,J,K \in \{1, \cdots,7\}$ and $a,b \in \{0,\cdots,7\}$.
We remark that octonions are nonassociative:
\begin{equation}
e^{I*} \Pro X  e^J \ = \  (e^{I*} X^*) \ ( X e^J) \ \not = \ e^{I*} e^J \quad,
\label{nonassoc}
\end{equation}
otherwise $S^7$ would be as simple as $S^3$ and many of the problems
we address and which remain would vanish.

\Eqn{S7} looks quite nonlinear, and is for that reason difficult to
generalize. The algebra becomes much more manageable if we first
represent $X$ by an uncomstrained octonion $\lambda$ via $X=\lambda /
|\lambda|$ and work with operators $K \ = \ \lambda J^*$ instead of
$J$. Then we obtain \cite{kalg}
\begin{eqnarray}
& K_a(z) K_b(w) \ = \ &{2\over z-w} \ \lambda_{[a} \ K_{b]}(w) \nonumber\\
& K_a(z) \lambda_b(w) \ = \ &{1\over z-w} \ (|\lambda|^2 \delta_{ab}
\ -\  \lambda_a \lambda_b)(w) \quad ,
\label{kalg}
\end{eqnarray}
which means that the structure functions are linear in the coordinate
$\lambda$. If we define fermionic generators $F^\alpha$
for $ \alpha \in \{0, \cdots ,7\}$ if N=8, or
$ \alpha \in \{1, \cdots ,N\}$ if $N<8$, this almost holds for the
structure functions of the entire superconformal algebra:
\begin{eqnarray}
& K_a(z) K_b(w) \ = \ &{2\over z-w} \ \lambda_{[a} \ K_{b]}(w) \nonumber\\
& K_a(z) F^\alpha(w) \ = \ &{1\over z-w} \ \sigma^{\alpha\beta}_{ab} \
				K_b(w) \ \partial \theta^\beta(w) \nonumber\\
&  F^\alpha(z) F^\beta(w) \ = \ &{-1\over (z-w)^2} \
	\Big(\ \lambda_{a}\ \sigma^{\alpha\beta}_{ab}\ K_{b}(z)\
	 \lambda_{a}\ \sigma^{\alpha\beta}_{ab}\ K_{b}(w) \ \Big) +\nonumber\\
& 	 &+\ {2\over z-w} \ \bigg( \delta^{\alpha\beta} \ \Big(
		|\lambda|^2 L - \partial \lambda_a K_a - \partial
		\theta^\alpha F^\alpha \Big) \ + \ 2 \partial \theta^{(\alpha}
		F^{\beta)}  \bigg) \quad .
\label{kfalg}
\end{eqnarray}
The structure functions in \eqn{kfalg} depend not only on the coordinates
of the seven-sphere, but also on their superpartners $\theta^\mu, \mu \in
\{0,\cdots,7\}$. It should be stressed that both \eqn{kalg} and
\eqn{kfalg} are computed as classical algebras, i.e. neglecting double
commutators. We should add the action of the generators on the coordinates:
\begin{eqnarray}
& K_a(z) \lambda_b(w) \ = \ &{1\over z-w} \ (|\lambda|^2 \delta_{ab}
\ -\  \lambda_a \lambda_b)(w)					 \nonumber\\
&  F^\alpha(z) \lambda_b(w) \ = \ &{-1\over (z-w)} \
	 \lambda_{a}\ (\sigma^{\alpha}\bar\sigma^{\beta})_{ab}
		\ \partial \theta^{\beta}(w) 			\nonumber\\
&  F^\alpha(z) \theta^\beta(w) \ = \ &{1\over z-w} \ \delta^{\alpha\beta}\
		|\lambda|^2(w) \quad.
\label{kfcoor}
\end{eqnarray}
A representation is furnished by the explicit example
\begin{eqnarray}
& K_a \ = \ &\lambda_b\lambda_b w_a \ -\ \lambda_b w_b \lambda_a
	\ - \ \half \sigma^{\mu\nu}_{ab}
	\lambda_b \psi^\mu \psi^\nu				 \nonumber\\
&  F^\alpha \ = \ &|\lambda|^2 p^\alpha \ + \
		\lambda_{a}\ (\sigma^{\alpha}\bar\sigma^{\beta})_{ab}\ w_b
		\ \partial \theta^{\beta}  \  		\nonumber\\
&		&+ \ 2 \ \psi^\alpha\ \psi^\beta\ \partial \theta^{\beta}
		\ + \ \partial x_a  (\sigma^{\alpha}\bar\sigma^{\mu})_{ab}
		\ \lambda_b \ \psi^\mu	\nonumber\\
& L \ = \ &\half\partial \lambda_a w_a \ - \ \half \lambda_a \partial w_a \
	\ + \ \half \partial x_a \partial x_a \ - \
	\half \psi_a \partial \psi_a  \quad .
\label{psixrep}
\end{eqnarray}
We may now rewrite the complete algebra in terms of the currents
$J^I(z)$ and $G_a(z) \ = \ |\lambda|^{-1} [X \ F \ e_{a *}]$.
We note that $G_a(z)$ is reducible for $N<8$.
\begin{eqnarray}
&J^I(z)G_a(w)\ =\ &{1\over z-w}\ \Bigg(
	(\bar\sigma^\alpha \sigma^I X)_a \ (X\sigma^\alpha G)
	\  						\nonumber\\
&	&+ \ 4 \ ( \delta^{IK}_{\alpha\beta} \ + \ \delta^{IJ} T^{IJK}_X \ )
	J^K \ (\bar\sigma^\alpha X)_a \ {\partial \theta^\beta \over |\lambda|}
	\Bigg)(w)					\nonumber\\
&G_a(z)G_b(w)\ =\ &{-1\over(z-w)^2}\ \Bigg(\
	(\bar\sigma^\alpha X)_a \ (\bar\sigma^\beta X)_b \
	\bigg( 	2 \delta^{0I}_{\alpha\beta} -
		\delta^{JK}_{\alpha\beta}T^{IJK}_X \bigg) \ J^I(z) \
							\nonumber\\
&	&\phantom{mmmmmmmmm}+ \ (z \longleftrightarrow w)\ \Bigg)
							\nonumber\\
&	&+ \ {2\over z-w}\  \ \Bigg(\
	(\bar\sigma^\alpha X)_{(a} \ (\bar\sigma^\beta \partial X)_{b)}\
	\bigg( 	2 \delta^{0I}_{\alpha\beta} -
		\delta^{JK}_{\alpha\beta}T^{IJK}_X \bigg) \ J^I(z) \
							\nonumber\\
&	&+ \ (\bar\sigma^\gamma X)_{(a} \
	(\bar\sigma^{\beta\delta}\sigma^\gamma)_{b)}\
	{\partial \theta^\delta \over |\lambda|} \ (X\sigma^\beta G)
							\nonumber\\
&	&+ \ (\bar\sigma^\alpha X)_a \ (\bar\sigma^\alpha X)_b
	\bigg( L \ + \ (X\sigma^I \partial X) J^I \bigg)
	\Bigg)(w)\qquad .
\label{jgalg}
\end{eqnarray}
The basic structure of the algebra becomes clear if we set $X=1$ and
$\theta = 0$, $\partial \theta = 0$:
\begin{eqnarray}
&J^I(z)G_a(w)\ \bigg|_{X=1,\theta = 0} \ =\
	&{1\over z-w}\ [ e^*_a  e^{I*}  e_\alpha ] \ G_\alpha(w)
	\  						\nonumber\\
&G_a(z)G_b(w)\ \bigg|_{X=1,\theta = 0} \ =\ & \delta^\alpha_a \delta^\beta_b
	\bigg\{ {1\over(z-w)^2}\  [ e^*_a  e_\beta^*  e^{I*}  ] \
	 ( J^I(z) \ + \ J^I(w)  ) \ 			\nonumber\\
&	&+ \  {2\over z-w}\  \delta^{\alpha\beta}
	\ L(w) \  \bigg\}
							\nonumber\\
&J^I(z)\psi^\mu(w)\ \bigg|_{X=1,\theta = 0} \ =\ &
	{1\over z-w}\ [ e^\mu  e^{\nu *}  e_{I*}  ] \ \psi^\nu
					 		\nonumber\\
&G_a(z)\psi^\mu(w)\ \bigg|_{X=1,\theta = 0} \ =\ &
	{1\over z-w}\ \delta^\alpha_a \ [ e^*_\alpha  e^{\mu}  e_b ]\
	\partial x_b (w) 				\nonumber\\
&G_a(z) x_b(w) \bigg|_{X=1,\theta = 0} \ = \ &
	{1\over z-w}\  \delta^\alpha_a \
	[ e_\alpha  e_b^* e^{\mu *}  ]\ \psi^\mu  \qquad ,
\label{shortalg}
\end{eqnarray}
which is an algebra very similar to the nonassociative construction of
Defever, Troost et.al.\cite{supernonasso}.
We interprete \eqn{shortalg} as a $N>4$ superconformal algebra
``at the north pole of $S^7$''. Almost all the complications evident in
\eqn{jgalg} are due to the nonassociativity of octonions forcing us off
the north pole.

The quantum versions of the above algebras are straightforward but
quite tedious to elaborate.
For $N<8$, field-dependent central terms arise. It is
easy to arrive at a quantum N=8 algebra
with only C-number central terms: one simply leaves out all the
$\psi$- and $x$-dependence in \eqn{psixrep}. It is more difficult,
but possible for N=8, to achieve that also in the general case.
For a thorough discussion of the various quantum versions
of the $N=8$ algebra we recommend H. Samtleben's discussion
\cite{samtleben}.

\section{S{\bf O}(1,9) and N=1 Supersymmetric Maxwell-Theory in D=10}

The off-shell supersymmetry transformations of a U(1) gauge theory
in D=(1,3) and D=(1,5) may be put in the form \cite{hypersusy}

\begin{eqnarray}
&\delta_\epsilon A_M  \ =\
	&2 i \ [ \epsilon_{\dot\alpha}^* \bar \sigma_M^{\dot\alpha \alpha}
		\psi_\alpha ]
							\nonumber\\
&\delta_\epsilon D^I \ =\
	&2 i \ [ \epsilon_{\dot\alpha}^* \bar \sigma_M^{\dot\alpha \alpha}
		\sigma^I{}_\alpha{}^\beta \partial^M \psi_\beta]
				 			\nonumber\\
&\delta_\epsilon \psi_\alpha\ = \
	& \partial_M A_N  \sigma^{MN}{}_\alpha{}^\beta \epsilon_\beta
 		\ - \ D^I \sigma^I{}_\alpha{}^\beta  \epsilon_\beta \qquad,
							\label{susyalg}
\end{eqnarray}

where $\alpha, \dot\alpha, \beta \in \{1,2\}$,
$M \in \{0, \cdots, D-1\}$,  $I \in \{1, \cdots, D-3\}$,
the spinor components $\epsilon_{\alpha},\psi_\alpha$ take
values in $\bf C$ or $\bf H$, and the bracket $[...]$ takes
the real part of a (hyper)complex expression.
The various sigma-matrices are defined as follows:

\begin{eqnarray}
&[\psi^{\alpha}\ \sigma^M{}_{\alpha \dot\alpha}
		\ \chi^{\dot\alpha} ]
	\ = \
	&\ \sqrt{2} \delta^M_{+} [\epsilon^*_1 \lambda_1] \
	-\ \sqrt{2} \delta^M_{-} [\epsilon^*_2 \lambda_2] \
	+ \  [\epsilon^*_1 e^\mu \lambda_2]\
	+ \  [\epsilon^*_2 e^{\mu *} \lambda_1]\
							\nonumber\\
&[\epsilon_{\dot\alpha}^*\ \bar \sigma^M{}^{\dot\alpha \alpha}
		\ \lambda_\alpha ]
	\ = \
	&-\ \sqrt{2} \delta^M_{-} [\epsilon^*_1 \lambda_1] \
	+\ \sqrt{2} \delta^M_{+} [\epsilon^*_2 \lambda_2] \
	- \  [\epsilon^*_1 e^\mu \lambda_2]\
	- \  [\epsilon^*_2 e^{\mu *} \lambda_1]\
							\nonumber\\
&[\psi^{\alpha} \ \sigma^I{}_{\alpha}{}^{\beta}\
		\lambda_{\beta} ]
	\ = \
	&\ [\psi^1 \lambda_1 e^{I*}] +  [\psi^2 \lambda_2 e^{I*}]
							\nonumber\\
&[\lambda^*_{\dot\alpha}\ \sigma^I{}^{\dot\alpha}{}_{\dot\beta}
		 \ \psi^*{}^{\dot\beta}]
	\ = \
	&\ [\lambda^{1*} \psi_1^* e^{I*}] +  [\lambda^{2*} \psi_2^* e^{I*}]
	\qquad ,
							\label{sigmamat}
\end{eqnarray}

where $\mu$ is a SO(D-2) vector index, i.e.
$\mu \in \{1,\cdots, D-2\}$,
and as usual $\sigma^{MN} \ = \ \sigma^{[M} \bar\sigma^{N]}$.
It is useful to introduce the real spinor
$\psi_A = \psi_{1,a}, \psi_{2,\dot a}$ in terms of two
SO(D-2)-spinors with $\psi_1 = \psi_{1,a} e_a$,
$\psi_2 = \psi_{2,\dot a}$, $a, \dot a  \in \{0,\cdots, D-3\}$.
Then the sigma-matrices have the following form:

\begin{eqnarray}
& \sigma^M_{AB} \ = \
	& \left( \begin{array}{cc}
		\sqrt{2} \delta^M_{+} \delta_{ab} & \sigma^\mu_{a\dot b} \\
		\bar\sigma^\mu_{\dot a b} & -\sqrt{2} \delta^M_{-}
		\delta_{\dot a \dot b}
		\end{array}
	  \right)
							\nonumber\\
& \bar\sigma^M{}^{AB} \ = \
	& \left( \begin{array}{cc}
		-\sqrt{2} \delta^M_{-} \delta_{ab} & -\sigma^\mu_{a\dot b} \\
		-\bar\sigma^\mu_{\dot a b} & \sqrt{2} \delta^M_{+}
		\delta_{\dot a \dot b}
		\end{array}
	  \right)
							\nonumber\\
& \sigma^I{}_A{}^B \ = \ \bar\sigma^I{}^A{}_B
	& \left( \begin{array}{cc}
		\sigma^I_{a b}& 0  \\
		0  & \sigma^I_{\dot a \dot b}
		\end{array}
	  \right)			\quad .
							\label{realsig}
\end{eqnarray}

The main properties of these sigma-matrices are
$\sigma^{(M} \bar\sigma^{N)} = - \eta^{MN}$ and $\sigma^M \bar\sigma^I =
\sigma^I \sigma_M$. The latter equation may be understood by
looking at \eqn{sigmamat}: $\sigma^M$ multiplies a spinor on the
left by a (hyper)complex number, while $\sigma^I$ does so on the
right. Both operations commute by the associativity of multiplication.
The reader may wonder about the matrices $\sigma^I$. They incorporate,
in a real basis, the (hyper)complex structure of spinors in
D=(1,3) and D=(1,5). One way to see why they are necessary is to
count sigma-matrices, for example in D=(1,5):
\begin{eqnarray}
\begin{array}{cccc}
\sigma^M_{(AB)} &: & 6 &\\
&&\\
\sigma^{MNP}_{[AB]} &: & 10 &\textrm{(selfdual in $MNP$)}\\
&&\\
(\sigma^M\bar\sigma^I)_{[AB]} &: & 3\cdot 6 & \\
&&\\
(\sigma^{MNP}\bar\sigma^I)_{(AB)} &: & 3 \cdot 10 &
\end{array}
						\label{sigmacount}
\end{eqnarray}
Without $\sigma^I$ we would not properly fill out the total of 64
matrices with 28 antisymmetric and 36 symmetric ones.
This counting makes it pretty clear why it should be hard to extend
this picture to D=(1,9). There we have:
\begin{eqnarray}
\begin{array}{cccc}
\sigma^M_{(AB)} &: & 10 &\\
&&\\
\sigma^{MNP}_{[AB]} &: & 120 &\\
&&\\
\sigma^{MNPQR}_{(AB)} &: & 126 & \textrm{(selfdual in $MNPQR$)} \quad .
\end{array}
						\label{sigma10count}
\end{eqnarray}
We obtain the full $16 \cdot 16 = 256$ matrices without recourse
to some $\sigma^I$, which would not commute with $\sigma^M$
in any event, due to the nonassociativity of octonions.

In the following we will introduce a modification of SO(1,9), which we
will call S{\bf O}(1,9), where there does exist an octonionic structure
that commutes with the generators.
Using this modification in \eqn{susyalg}, we will be able to retain
the closure of the algebra on the bosonic fields
$F_{MN} = \partial_{[M}A_{N]}$ and $D^I$, but not on the fermions
$\lambda_\alpha$.

To start with, we introduce the algebra of \eqn{S7} once again,
this time in the guise of what we will call S{\bf O}(8):
\begin{eqnarray}
& \Sigma^{\mu\nu}_{ab} \ = \
	& [ (\lambda e^{[\mu} ) e^{\nu]*} \partial_\lambda^* ] \delta_{ab}
 	\ + \  [ (e_a^* X^*) ((X e^{[\mu} ) e^{\nu]*}) e_b ]
						\label{soct8}
\end{eqnarray}
are the generators for the spinor-representation, so that,
in octonionic form, with $\zeta = \zeta_a e_a$:
\begin{eqnarray}
& 	\Sigma^{\mu\nu} \zeta \ = \
	& 	  X^* (((X e^{[\mu} ) e^{\nu]*}) \zeta)
						\nonumber\\
& 	\Sigma^{\mu\nu} X \ = \ & (X e^{[\mu} ) e^{\nu]*}
						\label{so8spinor}
\end{eqnarray}
The commutators are readily computed:
\begin{eqnarray}
& [ \Sigma^{\mu\nu} , \Sigma^{\rho\sigma} ]_{ab} \ = \
	& - 8 \delta^{[\mu}{}_{[\rho} \Sigma^{\nu]}{}_{\sigma]}{}_{ab}
	\quad ,
						\label{so8comm}
\end{eqnarray}
i.e. we obtain the usual SO(8) commutation relations.
However, we clearly are performing nothing but $S^7$-transformations,
and hence S{\bf O}(8) ought to be 7-dimensional. This is true, since the
projection operator
\begin{eqnarray}
& P^{\mu\nu}_{\rho\sigma} \ = \ &- {1\over 8}
	[ (e^{[\mu} \Pro X e^{\nu]*}) ( e^{[\rho} \Pro X e^{\sigma]*}) ]
						\label{projector}
\end{eqnarray}
acts trivially on $\Sigma^{\mu\nu}$ and has dimension 7:
\begin{eqnarray}
& P^{\mu\nu}_{\rho\sigma} \ \Sigma^{\rho\sigma}_{ab}
	\ = \ & \Sigma^{\mu\nu}_{ab}
						\nonumber\\
&  P^{\mu\nu}_{\mu\nu} \ = \ & 7 \qquad .
						\label{proact}
\end{eqnarray}
We have, through the introduction of the $S^7$-coordinate $X$,
gained the room to introduce (an infinity of) commuting
$S^7$'s, of which we pick the following example:
\begin{eqnarray}
& 	\Sigma^{I}_{ab} \ = \
	&  	\delta_{ab} \delta^I  \ + \
		[ e_a^* X^* ( ( ( X e_b ) Y^* ) ( Y e^{I*} )) ]
						\nonumber\\
& 	\delta^{I} Y \ = \ & Y e^{I*} \qquad,
						\label{comms7}
\end{eqnarray}
where we have introduced yet another $S^7$-variable Y, which is not
strictly necessary. Setting $Y=1$ and $\delta^{I} Y = 0$
does not alter the picture qualitatively. That case, however, is a rewritten
version of eq. (2.11) in \cite{cedpreit}.
In any case we have now
\begin{eqnarray}
& \Sigma^{\mu\nu} \Sigma^{I} \ = \ & \Sigma^{I} \Sigma^{\mu\nu} \quad .
						\label{commut}
\end{eqnarray}
It remains to define the counterpart to $\sigma^\mu_{a\dot b}$:
\begin{eqnarray}
& \Sigma^{0}_{a b} \ = \
	& \delta_{ab}
							\nonumber\\
& \Sigma^{j}_{a b} \ = \
	& [ \lambda  e^{j} \partial_\lambda^* ] \delta_{ab} \ + \
	[(e^*_a X^*) (X e^j) e_b]
							\nonumber\\
& \Sigma^{\mu}_{a \dot b} \ = \
	& (\Sigma^{0}, \Sigma^{j})_{a \dot b}
							\nonumber\\
& \overline\Sigma^{\mu}_{\dot a b} \ = \
	& (\Sigma^{0}, -\Sigma^{j})_{\dot a b}
							\nonumber\\
& \Sigma^{I}_{a b} \ = \
	& [ \eta e^{I*} \partial_\eta^* ] \delta_{ab} \ + \
	 [ (e_a^* X^*) ( ( X e_b) Y^* ) ( Y e^{I*} ) ]   \quad,
						\label{bigsigma}
\end{eqnarray}
where $Y = \eta / |\eta|$.
In order to ensure the usual (anti)hermiticity properties, we
integrate \eqn{sigmamat} over $X$ and $Y$, with normalization
$\int dX  = 1$. Without these
integrations, identities such as
\begin{eqnarray}
&\int dX dY [\psi^{a*}\ (\Sigma^{[i}\Sigma^j\Sigma^{k]}){}_{ab}
		\ \chi^{b} ]
	\ = \ \int dX dY [\chi^{a*}\ (\Sigma^{[i}\Sigma^j\Sigma^{k]}){}_{ab}
		\ \psi^{b} ]
							\label{reverse}
\end{eqnarray}
would be incorrect.

Now we have all the ingredients we need to construct
S{\bf O}(1,9): we simply replace SO(8)-invariant matrices
in \eqn{realsig} by  S{\bf O}(8)-invariant differential operators, and
inner products get integrated over $X$ and $Y$. We will define the
S{\bf O}(1,9)-invariant sigma-operators in complete analogy with
\eqn{realsig} and denote them with $\Sigma^M$, $\bar\Sigma^M$ and
$\Sigma^I$.

The supersymmetry algebra then reads:
\begin{eqnarray}
&\delta_\epsilon A_M  \ =\
	&2 i \ \int dX dY \
		[ \epsilon_{\dot\alpha}^* \bar \Sigma_M^{\dot\alpha \alpha}
		\psi_\alpha ]
							\nonumber\\
&\delta_\epsilon D^I \ =\
	&2 i \ \int dX dY \
		[ \epsilon_{\dot\alpha}^* \bar \Sigma_M^{\dot\alpha \alpha}
		\Sigma^I{}_\alpha{}^\beta \partial^M \psi_\beta]
				 			\nonumber\\
&\delta_\epsilon \psi_\alpha\ = \
	& \partial_M A_N  \Sigma^{MN}{}_\alpha{}^\beta \epsilon_\beta
 		\ - \ D^I \Sigma^I{}_\alpha{}^\beta  \epsilon_\beta \qquad,
							\label{susy10}
\end{eqnarray}
By construction, it closes off-shell on the bosonic fields.
Of course, this is the easy part. The algebra does not close on
the fermions, since the required Fierz-identity
\begin{eqnarray}
& \int dX dY dX' dY' \left(
\bar\Sigma_N(X){}^{D (B}\	\Sigma^{MN}(X'){}_A{}^{C)}
\ + \
\bar\Sigma^{MI}(X,Y){}^{D (B}\  \Sigma^{I}(X',Y'){}_A{}^{C)} \right) &
				 			\nonumber\\
&\ = \ 	 \int dX \ \bar\Sigma_M(X){}^{BC} \delta_A^D &
							\label{nofierz}
\end{eqnarray}
fails to hold. By writing $\bar\Sigma_M(X){}^{BC}$ we wish to
remind the reader of the dependence on the various $S^7$-variables.
Let us, as an example, consider the sector
$\alpha=1,\beta=1,\gamma=1,\delta=1$. Then \eqn{nofierz} reduces to
\begin{eqnarray}
& \delta_{d(b} \delta_{c)a} - \int dX dY \ \Sigma^{I}(X,Y){}_{d (b}
 		\ \int dX' dY' \ \Sigma^{I}(X',Y'){}_{c)a }
\ = \ & \ \delta_{bc} \delta_{ad} \quad,
							\label{nofierz8}
\end{eqnarray}
and we compute
\begin{eqnarray}
&\int dX dY \ \Sigma^{I}(X,Y){}_{a b}  \ = \ & - \frac{1}{4} [e^I e_a^* e_b]
							\nonumber\\
&  \int dX dY \ \Sigma^{I}(X,Y){}_{d (b}
 		\int dX' dY' \ \Sigma^{I}(X',Y'){}_{c)a } \ = \
	& \frac{1}{16} ( \delta_{d(b} \delta_{c)a} - \delta_{bc}\delta_{ad} )
	\quad .
							\label{sigmai}
\end{eqnarray}
We obtain the correct tensor structure, but with an additional
coefficient $1/16$, which ruins the show. Curiously, we
also obtain
\begin{eqnarray}
&  \Sigma^{I}(X,Y){}_{d (b}  \Sigma^{I}(X,Y){}_{c)a } \ = \
	&  \delta_{d(b} \delta_{c)a} - \delta_{bc}\delta_{ad}
	\quad ,
							\label{sigmaixy}
\end{eqnarray}
which looks, up to the missing integrations, just like \eqn{nofierz8}.
Such tantalizing coincidences show up also in other sectors of \eqn{nofierz},
but we have not been able to use them to define a consistent off-shell
supersymmetry algebra in D=(1,9).

\newpage

\noindent
{\bf Acknowledgements}
\vspace{5mm}

\noindent
I thank N. Berkovits, L. Brink and M. Cederwall for
extensive discussions, as well as the organizers of the
workshop for an enjoyable meeting.

\end{document}